\documentclass[12pt]{article}
\usepackage{amsmath,amssymb,latexsym,float}
\usepackage{enumerate}

\numberwithin{equation}{section}

\newcommand{\R}{\text{\fontshape{n}\selectfont I\kern-.42exR}}

\newcommand{\1}{\text{\fontshape{n}\selectfont 1\kern-.56exl}}

\newcommand{\pslash}{\slash\!\!\!p}
\newcommand{\qslash}{\slash\!\!\!q}

\begin{document}
\title{
{\bf Creutz Fermions on an Orthogonal Lattice}
}

\author{Artan Bori\c{c}i\\
        {\normalsize\it University of Tirana}\\
        {\normalsize\it Department of Physics, Faculty of Natural Sciences}\\
        {\normalsize\it King Zog I Boulevard, Tirana, Albania}\\
        {\normalsize\it borici@fshn.edu.al}\\
}

\date{}
\maketitle

\vspace{2cm}
\begin{abstract}
In a recent paper, Creutz has given a new action describing two species of Dirac fermions with exact chiral symmetry on the lattice. This action depends on parameters which may be fixed at certain values in order to get the right continuum limit. In this letter, we elaborate more on this idea and present an action which is free of any other parameter except the fermion mass.
\end{abstract}

\vspace{14cm}
\pagebreak


\setcounter{section}{1}

Ever since the birth of lattice QCD it was understood that chiral symmetry cannot be easily realised on the lattice \cite{Wilson_1974}. The naive discretization of the Dirac operator on a regular hypercubic lattice, as given in the momentum space,
$$
D(p)=\sum_{\mu}~i\gamma_{\mu}~\sin~p_{\mu}\ ,
$$
possesses exactly 16 zeros in the 16 corners of the Brillouin zone. Since they differ only on their slope, one can interpret them as 16 degenerated Dirac species.

Later, Wilson lifted this degeneracy by defining a new operator, namely \cite{Wilson_Erice_1977},
$$
D(p)=\sum_{\mu}~i\gamma_{\mu}~\sin~p_{\mu}+\sum_{\mu}~(1-\cos~p_{\mu})\ .
$$
This way, one gets a single Dirac fermion on the lattice with the chiral symmetry explicitly broken.

Another approach, the staggered fermion approach, puts the four components of a Dirac spinor on different lattice sites \cite{KS_1975}. This reduces the degeneracy to four species.

Further research on chiral symmetric lattice fermions was discouraged by Nielsen and Ninomiya theorem, which states the impossibility of a single left handed chiral fermion on the lattice \cite{NN_1981}. However, as pointed out by Wilczek, the theorem does not prevent having a pair of Dirac fermions on the lattice \cite{Wilczek_1987}. So, he argues, if $f_1(p),f_2(p),f_3(p),f_4(p)$ are suitable functions of lattice momenta one can get a Dirac operator,
$$
D(p)=\sum_{\mu}~i\gamma_{\mu}~f_{\mu}(p)\ ,
$$
with only one doubler. His explicit construction,
\begin{eqnarray*}
f_4&=&\frac1a\left\{\sin~p_4a+\lambda\left(\sin^2~\frac{p_1a}2+\sin^2~\frac{p_2a}2+\sin^2~\frac{p_3a}2\right)\right\},~~~~~\lambda>1\\
f_j&=&\frac1a\sin~p_ja,~~~~~j=1,2,3\ ,
\end{eqnarray*}
exhibits two zeros: one at $(0,0,0,0)$ and the other at $(0,0,0,\pi)$. However, the loss of cubic symmetry introduces extra relevant terms in the Wilson plaquette action \cite{Wilczek_1987}.

Recently, motivated by the Dirac structure of graphene electrons in two dimensions, Creutz was able to elegantly generalise this structure to four dimensions, exactly what one needs in particle physics \cite{Creutz_07}. This generalisation inherits the gapless property of the electronic pair which comes from the fact that the electron and its doubler sit symmetrically with respect to the centre of the Brillouin zone. He starts by defining the left handed degrees of freedom in 4-momentum space:
$$
\begin{matrix}
z(p)=B&[4C-\cos p_1-\cos p_2-\cos p_3 -\cos p_4\\
&+i\sigma_1(\sin p_1+\sin p_2-\sin p_3-\sin p_4)\\
&+i\sigma_2(\sin p_1+\sin p_2-\sin p_3-\sin p_4)\\
&+i\sigma_3(\sin p_1+\sin p_2-\sin p_3-\sin p_4)]
\end{matrix}
$$
where $\sigma_k, k=1,2,3$ are the Pauli matrices and $B,C$ are real parameters. In order to get the Dirac fermion we bring this expression in the form:
$$
z(p)=B[4C-\sum_{\mu}\cos p_{\mu} + i~\sum_{k=1}^3\sigma_k s_k(p)]\ ,
$$
where $s_k(p)$ are defined through:
$$
\begin{matrix}
s_1(p)=\sin p_1+\sin p_2-\sin p_3-\sin p_4\\
s_2(p)=\sin p_1-\sin p_2-\sin p_3+\sin p_4\\
s_3(p)=\sin p_1-\sin p_2+\sin p_3-\sin p_4
\end{matrix}\ .
$$
Introducing right movers as $\bar z(p)$ we get the Dirac operator in the form:
$$
Z(p)=\begin{pmatrix} z(p) & \\ & \bar z(p) \end{pmatrix}\ .
$$
Using the Euclidean Dirac gamma matrices in the chiral representation
$$
\gamma_k=\begin{pmatrix} & -i\sigma_k \\ i\sigma_k & \end{pmatrix}\ ,~~~~~~
\gamma_4=\begin{pmatrix} & ~~~\1 \\ \1~~~ & \end{pmatrix}\ ,
$$
we get the momentum space Creutz-Dirac operator:
\begin{equation}\label{Creutz_operator}
aD_C(p)=iB\gamma_4(4C-\sum_{\mu}\cos ap_{\mu}) + iB~\sum_{k=1}^3\gamma_k s_k(ap)\ ,
\end{equation}
where we have restored the lattice spacing $a$. Creutz has shown that the zeros of this operator in the Brillouin zone are at $(\tilde p,\tilde p,\tilde p,\tilde p)$ and $(-\tilde p,-\tilde p,-\tilde p,-\tilde p)$, where $C=\cos \tilde p$. The reader can immediately see that the formal continuum limit, i.e. $a\rightarrow 0$, does not give the continuum Dirac operator $\pslash$. While this is not a problem, it is related to the fact that the Creutz action has no zero at the origin, making its continuum limit subtle. In order to get a hyper-cubic lattice of $p$-translations, Creutz shows that its parameters must satisfy $BS=C$, where $S=\sin \tilde p$.

From the discussion above it is clear that it is desirable to have an action with orthogonal p-axes which has a formal continuum limit. While these features may not be necessary \cite{Bedaque_et_al}, they serve us as a guiding principle to fix the parameter $C$. We shall show that such an action exists and has the form:
\begin{equation}\label{G_action}
D(p)=\sum_{\mu}~i\gamma_{\mu}~\sin~p_{\mu}+\sum_{\mu}~i\gamma'_{\mu}(\cos~p_{\mu}-1)\ ,
\end{equation}
where $\gamma'_{\mu},\mu=1,2,3,4$ is another set of Dirac gamma matrices, which are given as a linear combination of $\gamma_{\mu}$,
$$
\gamma'_{\mu}=\sum_{\nu}~\alpha_{\mu\nu}\gamma_{\nu}\ ,
$$
and $\alpha_{\mu\nu}$ are the coefficients of the orthogonal matrix $\alpha$. This action possesses one zero at $p=0$, the location of the rest depending on the details of $\gamma'$-matrices. In the rest of the paper we will show that this action can be derived from the Creutz action. In doing so we will find the concrete form of $\alpha$.

We start by dropping the overall parameter $B$ and setting again lattice spacing to one:
$$
D_C(p)=i\gamma_4(4C-\sum_{\mu}\cos ap_{\mu}) + i~\sum_{k=1}^3\gamma_k s_k(ap)\ .
$$
Translating momenta, $p_{\mu}=\tilde p+q_{\mu}$, we have:
$$
D_C(q)=i\gamma_4(4C-C\sum_{\mu}\cos q_{\mu}) + iC~\sum_{k=1}^3\gamma_k s_k(q)+iS\gamma_4\sum_{\mu}\sin q_{\mu}+iS~\sum_{k=1}^3\gamma_k c_k(q)\ ,
$$
where
$$
\begin{matrix}
c_1(q)=(\cos q_1-1)+(\cos q_2-1)-(\cos q_3-1)-(\cos q_4-1)\\
c_2(q)=(\cos q_1-1)-(\cos q_2-1)-(\cos q_3-1)+(\cos q_4-1)\\
c_3(q)=(\cos q_1-1)-(\cos q_2-1)+(\cos q_3-1)-(\cos q_4-1)
\end{matrix}\ .
$$
In order to simplify the action we assume $C+S=0$, which gives $C=1/\sqrt{2}$ conforming the $C>1/2$ constraint imposed by Creutz. Rescaling the action and defining:
\begin{eqnarray*}
s_4(q)&=&-\sin q_1-\sin q_2-\sin q_3-\sin q_4\\
c_4(q)&=&(\cos q_1-1)+(\cos q_2-1)+(\cos q_3-1)+(\cos q_4-1)\ ,
\end{eqnarray*}
the {\it translated} Creutz action takes the form:
$$
D_C(q)=\sum_{\mu}i\gamma_{\mu} s_{\mu}(q) + \sum_{\mu}i\gamma_{\mu} c_{\mu}(q)\ ,
$$
or in the scalar product notation, $(\gamma,x)=\sum_{\mu}\gamma_{\mu}x_{\mu}$, one has:
$$
D_C(q)=i(\gamma,s(q)+c(q))\ .
$$
Now, introducing the following orthogonal matrices:
$$
a:=\frac12
\begin{pmatrix}
 1 &  1 & -1 & -1 \\
 1 & -1 & -1 &  1 \\
 1 & -1 &  1 & -1 \\
-1 & -1 & -1 & -1 \\
\end{pmatrix}\ ,
~~~~~~b:=-\frac12
\begin{pmatrix}
 1 &  1 & -1 & -1 \\
 1 & -1 & -1 &  1 \\
 1 & -1 &  1 & -1 \\
 1 &  1 &  1 &  1 \\
\end{pmatrix}\ ,
$$
and noting that,
$$
s=2a\tilde s, ~~~~~~c=2b\tilde c\ ,
$$
where
$$
\tilde s=(~\sin q_1,~\sin q_2,~\sin q_3,~\sin q_4~)^T, ~~~~\tilde c=(\cos q_1 - 1,~\cos q_2 - 1,~\cos q_3 - 1,~\cos q_4 - 1~)^T\ ,
$$
then, the rescaled action by a factor of $2$ can be written in the form:
$$
D_C(q):=i(\gamma,a\tilde s(q)+b\tilde c(q))=i(a^T\gamma,\tilde s(q)+a^Tb\tilde c(q))\ .
$$
Denoting,
$$
\alpha:=a^Tb=\frac12
\begin{pmatrix}
-1 &  1 &  1 &  1 \\
 1 & -1 &  1 &  1 \\
 1 &  1 & -1 &  1 \\
 1 &  1 &  1 & -1 \\
\end{pmatrix}\ ,
$$
we get:
$$
D_C(q)=i(a^T\gamma,\tilde s(p)+\alpha\tilde c(q))\ .
$$
It is easy to show that $a^T\gamma$ are Dirac gamma matrices
\footnote{
$$
\{(a^T\gamma)_{\mu},(a^T\gamma)_{\nu}\}=\sum_{\rho,\sigma}a_{\rho\mu}a_{\sigma\nu}\{\gamma_{\rho},\gamma_{\sigma}\}=2\sum_{\rho}a_{\rho\mu}a_{\rho\nu}=2\delta_{\mu\nu}\ .
$$
}
, therefore the factor $a^T$ can be droped. This way, our final expression is:
\begin{eqnarray*}
D_C(q)&=&i(\gamma,\tilde s(q))+i(\gamma',\tilde c(q))\\
&=&\sum_{\mu}~i\gamma_{\mu}~\sin~q_{\mu}+\sum_{\mu}~i\gamma'_{\mu}(\cos~q_{\mu}-1)\ ,
\end{eqnarray*}
where $\gamma'=\alpha\gamma$ are again Dirac gamma matrices. This is exactly the action of eq. (\ref{G_action}). Noting that,
$$
\sum_{\mu}\gamma_{\mu}=\sum_{\mu}\gamma'_{\mu}\equiv2\Gamma\ ,
$$
we get another expression for the fermion action:
\begin{equation}\label{final_form}
D_C(q)=\sum_{\mu}~i\gamma_{\mu}~\sin~q_{\mu}+\sum_{\mu}~i\gamma'_{\mu}\cos~q_{\mu}-2i\Gamma\ .
\end{equation}
It is easy to see that the formal continuum limit of this action is:
$$
D_C(q)\rightarrow \qslash\ ,
$$
hence, this action has one zero at $q=0$. It is easily verified that the other zero is at $(\pi/2,\pi/2,\pi/2,\pi/2)$. Since we arrived at this action starting from the Creutz action with equivalent transformations, which conserve the number of zeros, then we may conclude that there are no more zeros. Hence, the action defined in eq. (\ref{final_form}) describes two species of chirally symmetric Dirac fermions. The slope at the second zero can be computed by expressing $q_{\mu}=\pi/2+k_{\mu}$ in eq. (\ref{final_form}):
$$
D_C(\pi/2+k)=\sum_{\mu}~i\gamma_{\mu}~(\cos~k_{\mu}-1)-\sum_{\mu}~i\gamma'_{\mu}\sin~k_{\mu}\rightarrow -\sum_{\mu}~i\gamma'_{\mu}k_{\mu}=-\sum_{\mu}\gamma_{\mu}(\alpha k)_{\mu}\ .
$$
Therefore, the slope at $(\pi/2,\pi/2,\pi/2,\pi/2)$ is the matrix $-\alpha$.

In order to write down the action in the position space, we use eq. (\ref{final_form}). Adding a mass term and multiplying both sides by $i$ we get:
$$
iD_C(q)=im+2\Gamma-\frac12\sum_{\mu}\left[\left(\gamma'_{\mu}-i\gamma_{\mu}\right)e^{iq_{\mu}}+\left(\gamma'_{\mu}+i\gamma_{\mu}\right)e^{-iq_{\mu}}\right]\ .
$$
Hence, the position space Dirac operator involves the following three terms:
\begin{itemize}
\item[] the diagonal term, $im+2\Gamma$;
\item[] the forward hopping term in the $\mu$ direction, $-\frac12(\gamma'_{\mu}-i\gamma_{\mu})$;
\item[] the backward hopping term in the $\mu$ direction, $-\frac12(\gamma'_{\mu}+i\gamma_{\mu})$.
\end{itemize}
Finally, the gauge fields can be introduced in the standard way, the resulting action being gauge invariant.

In conclusion, based on the original action of Creutz, we have given a minimally doubled and chirally symmetric fermion action. The formal continuum limit served us as a guiding principle to fix its parameters. However, the parameters have geometric significance and can be fixed at other values in order to maximize the symmetries of the action \cite{Creutz_07,Bedaque_et_al}. As was shown in detail at references \cite{Bedaque_et_al,Creutz_08}, the lack of hyper-cubic symmetry may introduce relevant terms in the presence of interactions. Thus, as we already know, chiral symmetry on the lattice comes with a high price. Whether the price of the minimally doubled action is higher or lower than the price of other actions, this is a question that can be settled by direct simulations \cite{Cichy_et_al}.

\section*{Acknowledgements}

The author would like to thank Mike Creutz for discussions related to his action.


\begin{thebibliography}{10}

\bibitem{Wilson_1974} K. G. Wilson, Phys.Rev.D10:2445-2459,1974.

\bibitem{Wilson_Erice_1977} K. G. Wilson in  {\it New Phenomena In Subnuclear Physics}, ed. A. Zichichi, Plenum Press, New York, 1977.

\bibitem{KS_1975} J.B. Kogut, L. Susskind, Phys.Rev.D11:395,1975.

\bibitem{NN_1981} H.B. Nielsen, M. Ninomiya, Nucl.Phys.B185:20,1981, Erratum-ibid.B195:541,1982.

\bibitem{Wilczek_1987} F. Wilczek, Phys.Rev.Lett.59:2397,1987.

\bibitem{Creutz_07} M. Creutz, JHEP 0804, 017 (2008), ArXiv:0712.1201 [hep-lat].

\bibitem{Bedaque_et_al} P. F. Bedaque, M. I. Buchoff, B. C. Tiburzi and A. Walker-Loud, Phys. Lett. B 662, 449 (2008), ArXiv:0801.3361 [hep-lat]. P. F. Bedaque, M. I. Buchoff, B. C. Tiburzi and A. Walker-Loud, Phys. Rev. D 78, 017502 (2008) ArXiv:0804.1145 [hep-lat].

\bibitem{Creutz_08} M. Creutz, {\it Local chiral fermions}, ArXiv:0808.0014 [hep-lat].

\bibitem{Cichy_et_al} K. Cichy, J. Gonzalez Lopez, K. Jansen, A. Kujawa and A. Shindler, Nucl. Phys. B 800, 94 (2008), ArXiv:0802.3637 [hep-lat].



\end{thebibliography}
\end{document}